\documentclass[%
preprint,
superscriptaddress,
 amsmath,amssymb,
 aps,
]{revtex4-1}

\usepackage{graphicx}
\usepackage{dcolumn}
\usepackage{bm}
\usepackage{amssymb}
\usepackage{caption}
\usepackage{subcaption}
\usepackage{amsmath}
\usepackage[T1]{fontenc}
\usepackage[utf8]{inputenc}
\usepackage[usenames, dvipsnames]{color}
 \usepackage{setspace}
 \usepackage{float}
 \usepackage{array}
 \usepackage{multirow}
\newcolumntype{P}[1]{>{\centering\arraybackslash}p{#1}}
\newcolumntype{M}[1]{>{\centering\arraybackslash}m{#1}}

\begin{document}


\title{A numerical lift force analysis on the inertial migration of a deformable droplet in steady and oscillatory microchannel flows at different Capillary numbers and oscillation frequencies}

\def\correspondingauthor{\footnote{Corresponding author: alafzi@purdue.edu}}

\author{Ali Lafzi \correspondingauthor{}}
 \affiliation{Department of Agricultural and Biological Engineering, Purdue University, West Lafayette, Indiana 47907, USA}

\author{Sadegh Dabiri}
 \affiliation{Department of Agricultural and Biological Engineering, Purdue University, West Lafayette, Indiana 47907, USA}
 \affiliation{School of Mechanical Engineering, Purdue University, West Lafayette, Indiana 47907, USA}




\begin{abstract}
Inertial migration of deformable particles has become appealing in recent years due to its numerous applications in microfluidics and biomedicine. The physics underlying the motion of such particles is contingent upon the presence of lift forces in microchannels. This importance initiated several works to analyze and quantify such forces acting on particles. However, since most of such attempts have focused on solid and non-deformable particles, we extend the lift force analysis for the case of deformable droplets and study the effects of Capillary numbers on their dynamics in this paper. Furthermore, utilizing oscillatory flows as an alternative for steady currents within the microchannels has been proved to be beneficial by introducing new equilibrium positions for the particles. Therefore, the present analysis includes the oscillatory regimes and identifies the effects of oscillation frequencies on lift forces as well. We then propose an expression that mimics the lift force behavior in oscillatory flows accurately. Finally, we introduce a procedure to derive and predict a simple expression for the steady and averaged oscillatory lift for any given combination of Capillary number and oscillation frequency within a continuous range.
\end{abstract}

\maketitle


\section{\label{sec1}Introduction}
Inertial migration of particles in microchannels has caught extensive attention in the last two decades due to its numerous applications in cell sorting, fractionation, filtration, and separation in many clinical practices \cite{gossett2010label, toner2005blood, gossett2012hydrodynamic, karimi2013hydrodynamic}. The presence of lift forces acting on these particles is the chief reason for observing the underlying physical phenomena in microfluidic systems \cite{di2009inertial, martel2014inertial, bazaz2020computational, connolly2020mechanical, stoecklein2018nonlinear}. The importance of these forces has motivated many researchers to analyze or measure them within the microchannel. \citeauthor{di2009particle} have derived the inertial lift on particles and studied the effects of channel Reynolds number and particle size on it; they have shown that by increasing Reynolds, the magnitude of lift coefficient decreases near the wall and increases near the channel center. Also, the particle equilibrium positions shift toward the center as its size increases and its rotational motion is not a key component of the inertial lift \cite{di2009particle}. Using lift force profiles, \citeauthor{prohm2014feedback} have investigated and categorized the particle focusing points and demonstrated that the stable fix points lie on either the diagonal or main axes of the channel cross-section \cite{prohm2014feedback}. A fast numerical algorithm combined with machine learning techniques has been proposed to predict the inertial lift distribution acting on solid particles over a wide range of operating parameters in straight microchannels with three types of geometries by specifying the cross-sectional shape, Reynolds number, and particle size \cite{su2021machine}.

Furthermore, there have been attempts to derive analytical relationships for the observed behaviors. A simple formula using data fitting and least square was obtained to investigate the relationship between the lift and particle size and Reynolds number; according to the proposed criterion, particle focusing does not occur for too small particles or too low Reynolds numbers \cite{wang2017analysis}. \citeauthor{asmolov2018inertial} illustrated that the velocity of finite-size particles near the channel wall is different from that in the undisturbed flow and then reported a generalized expression for the lift force at Re $\leq$ 20 \cite{asmolov2018inertial}. Another study has proposed a generalized formula for the inertial lift acting on a sphere that consists of 4 terms: wall-induced lift, shear-gradient-induced lift, slip-shear lift, and correction of the shear-gradient lift; the authors have further confirmed that wall and shear-gradient are the main features of the lift \cite{liu2016generalized}. Moreover, there are examples of works concentrating on the effect of particle shape. For instance, \citeauthor{zastawny2012derivation} presented the great influence of shape both by changing the experienced values of forces and torques and modifying the Reynolds at which the transition to unsteady flow happens \cite{zastawny2012derivation}. Further extension on previous theories and analytical works resulted in an analytical expression capturing the weak, inertial lift on an arbitrarily-shaped particle moving near a wall \cite{mannan2020weak}.

Most of the studies on lift forces in the microchannels have focused on solid particles or non-deformable objects and have analyzed the effect of parameters such as channel Reynolds, particle size, etc. Therefore, there are very few examples presenting the whole lift force profiles acting on deformable particles such as droplets and bubbles and studying the effect of their corresponding parameters like Capillary number on the force values. For example, \citeauthor{chen2014inertial} have extensively studied the inertial migration of a deformable droplet in a rectangular microchannel, but their presented lift force profile only considers one value for particle Weber number (a measure for particle deformability) \cite{chen2014inertial}. \citeauthor{rivero2018bubble} have divided the underlying physics into different regimes. In the pure inertial regime, they have plotted the inertial lift on a rigid bubble at different Reynolds numbers, and in the pure Capillary regime where the inertial effects are absent, a lift profile is presented for different Capillary numbers \cite{rivero2018bubble}. However, their work lacks a similar profile visualizing the total lift force in the most general nonlinear inertial-capillary regime.

The obtained lift force profiles are mainly the result of either some experimental measurements \cite{di2009particle, stan2011sheathless, zhou2013fundamentals} or applying a feedback control in the numerical code to fix the position of particle \cite{raffiee2019numerical}, capsule \cite{schaaf2017inertial}, or drop \cite{chen2014inertial}. Nevertheless, in this paper, we present a method for lift force calculation at different Capillary numbers that solely depends on the trajectory of the drop. In addition, the importance of exploiting an oscillatory flow in the microchannel for working with sub-micron particles \cite{mutlu2018oscillatory}, having direct control over their focal points and tuning them depending on the flow oscillation frequency \cite{lafzi2020inertial}, and more effective separation and sorting strategies \cite{schonewill2008oscillatory} has already been presented in the literature. Thus, we will expand our lift force analysis to include both steady and oscillatory regimes at various Capillary numbers, where the latter is completely missing in the literature. We will then try to fit analytical expressions to the obtained lift profiles for different cases and present a scheme to predict this expression over a continuous range of input parameters.

\section{\label{sec:level1}Methodology\protect\\}
\begin{figure}[h]
  \centering
  \includegraphics[height=3in]{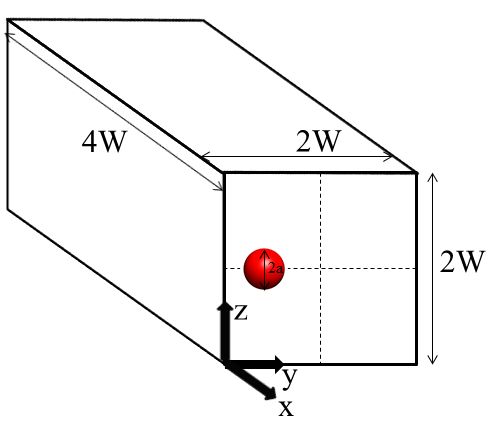}
   \caption{Schematic of the problem setup}
   \label{geometry}
\end{figure}

A single droplet with density and viscosity ratios of one is placed in a laminar flow of an incompressible Newtonian fluid in a microchannel as illustrated in Fig. \ref{geometry}. The drop dynamics is simulated using Front-tracking method \cite{unverdi1992front} as elaborated in detail in our previous work \cite{lafzi2021dynamics}. The pressure gradient in the $x$ direction has a constant magnitude of $P_0$ for the steady flow and a varying strength of  ${P_0}cos({\omega}t)$ for the oscillatory flow. The periodic boundary condition is applied in the $x$ direction, and the no-slip condition is applied on the walls in the $y$ and $z$ directions. The axes of symmetry have been avoided. Parameters $W$ and $U_c$ (maximum velocity of the steady case) are used as the characteristic length and velocity, respectively. In other words, $x^*={\frac{x}{W}}$, $u^*={\frac{u}{U_c}}$, $t^*={\frac{t}{{\frac{W}{U_c}}}}$, $P^*={\frac{P}{\mu{\frac{U_c}{W}}}}$, $T^*={\frac{T}{{\frac{W}{U_c}}}}$ (where $T$ is the period), and $\omega^*={\frac{2\pi}{T^*}}$. Three dimensionless parameters describe the dynamics of the drop: (i) Reynolds number, $Re={\frac{{\rho}U_c2W}{\mu}}$, where $\rho$ and $\mu$ are the density and viscosity, respectively, (ii) Capillary number, $Ca={\frac{\mu U_c}{\gamma}}$, in which $\gamma$ is the surface tension, and (iii) the dimensionless oscillation frequency ($\omega^*$). The drop has a constant size of $\frac{a}{W}=0.3$ with a spherical initial shape, and $Re=10$ in our entire study. The numerical grid is generated using $196\times 114 \times 114$ cells in the $x$, $y$, and $z$ directions, respectively, and with 29578 triangular elements for the discretization of the drop.

The active dominant forces on the migrating drop in the wall-normal direction are the inertial and deformation-induced lift and lateral drag forces. The direction of the inertial lift at drop locations far from the wall is towards it, and the deformation lift pushes the drop towards the channel center \cite{zhang2016fundamentals}. The direction of the lateral drag is the negative of its migration velocity sign, assuming that the carrier fluid is stationary in the wall-normal direction \cite{zhang2016fundamentals}. Therefore, if we assume the positive direction to be the one from the center to the wall, the force balance on the drop according to Newton's second law is the following:
\begin{equation}\label{force balance}
F_{total}=F_{inertial}-F_{deformation}-F_{drag},
\end{equation}
The drag force is computed based on its definition \cite{ishii1979drag}:
\begin{equation}\label{drag force}
F_{drag}=\frac{1}{2}C_D\rho v_r|v_r|A_d,
\end{equation}
\begin{equation}\label{drag coeff}
C_D=\frac{24}{Re_{rel}}(1+0.1Re_{rel}^{0.75}),
\end{equation}
\begin{equation}\label{Re_rel}
Re_{rel}=Re\frac{v_r}{U_c}\frac{a}{W},
\end{equation}
Where $C_D$ is the drag coefficient, $v_r$ is the relative velocity between the drop and the fluid (which is essentially its migration velocity), and $A_d$ is the frontal projected area of the drop. Equation \ref{drag coeff} is consistent with the findings of \cite{snyder2007statistical, ishii1979drag, zhou2020analyses, kelbaliyev2007development, salibindla2020lift} and those of \cite{ceylan2001new} at a viscosity ratio of one. Although eq. \ref{drag coeff} is derived for steady flows, researchers have shown that the drag coefficient in unsteady flows depends heavily on an unsteady parameter that includes the density ratio \cite{shao2017detailed, aggarwal1995review}. Since the density ratio in the present study is one, the aforementioned unsteady parameter becomes zero, and hence, $C_D$ for unsteady flows (including oscillatory cases) can be approximated as the one for steady flows using this equation. The parameter $A_d$ is calculated based on the projected area of the drop on a plane having a normal vector parallel to its migration velocity. Thus, the value of this parameter varies at different instances. This procedure leads to a more precise computation of the drag force.

Considering a viscosity ratio of one, the deformation-induced lift for a drop that has a distance higher than its diameter from the closest wall leads to the following compact form \cite{chan1979motion, stan2013magnitude}:
\begin{equation}\label{deformation force}
F_{deformation} = 75.4Ca_p\mu V_{avg}a(\frac{a}{W})^2\frac{d}{W},
\end{equation}
Where $Ca_p=Ca{\frac{a}{W}}$ is the drop capillary number, $V_{avg}$ is the average velocity of the carrier fluid across the channel, and $d$ is the distance of the drop from the channel center. The linear dependency of this force with respect to the distance $d$ in the specified region is also confirmed in \cite{rivero2018bubble}.

The lift force analysis in this work is solely based on the drop trajectory. Therefore, to get a lift profile that spans a wide range of $d$, the drop is released from two different initial locations: \begin{itemize}
    \item $y^*=0.46$ and $z^*=1$ (the upper release)
    \item $y^*=0.98$ and $z^*=1$ (the lower release)
\end{itemize}  
The upper release is chosen such that the whole range of studied $d$ falls within the validity domain of the deformation force equation (eq. \ref{deformation force}). This enables us to plug eq. \ref{deformation force} into the force balance equation (eq. \ref{force balance}) to get the inertial force once the total force is calculated as elaborated below. We will compare the inertial force at different $Ca$ values for the steady flows in the results section. The lower release initial location is slightly off from the channel center since it is also an equilibrium point, and if a drop is placed there, it does not move at all \cite{chen2014inertial}. The initial $z$ component for both releases is on the main axis for faster convergence since the drop eventually focuses on the main axes according to our previous work \cite{lafzi2021dynamics}. These different initial locations do not alter the drop equilibrium position \cite{pan2016motion, lan2012numerical, chaudhury2016droplet, razi2017direct}. The results of each parameter computation for both releases will be combined to reflect its overall behavior within the channel cross-section.

The migration velocity and acceleration of the drop is calculated by taking the first and second temporal derivatives from its trajectory numerically. Since time-step varies throughout the simulations to keep the Courant–Friedrichs–Lewy number at 0.9, the following equations are used to obtain the corresponding derivatives \cite{sundqvist1970simple}:
\begin{equation}\label{first derivative}
 v_r=\dot{d}_i \approx \frac{-h_i}{(h_{i-1})(h_i + h_{i-1})}d_{i-1} + \frac{h_i - h_{i-1}}{h_i h_{i-1}}d_i + \frac{h_{i-1}}{(h_i)(h_i + h_{i-1})}d_{i+1},
\end{equation}
\begin{equation}\label{second derivative}
 \frac{dv_r}{dt}=\ddot{d}_i \approx \frac{2\left[ d_{i+1} + \frac{h_i}{h_{i-1}}d_{i-1} -(1 + {\frac{h_i}{h_{i-1}}} d_i) \right]}{h_i h_{i-1} (1 + {\frac{h_i}{h_{i-1}}})},
\end{equation}
In which $h_i = t_{i+1} - t_i$, $h_{i-1} = t_i - t_{i-1}$, and $d_i$ and $t_i$ denote the distance from center and time at the current step. Both non-uniform finite difference schemes have a second-order accuracy. 

\begin{figure}[h]
  \centering
  \includegraphics[width=4.3in]{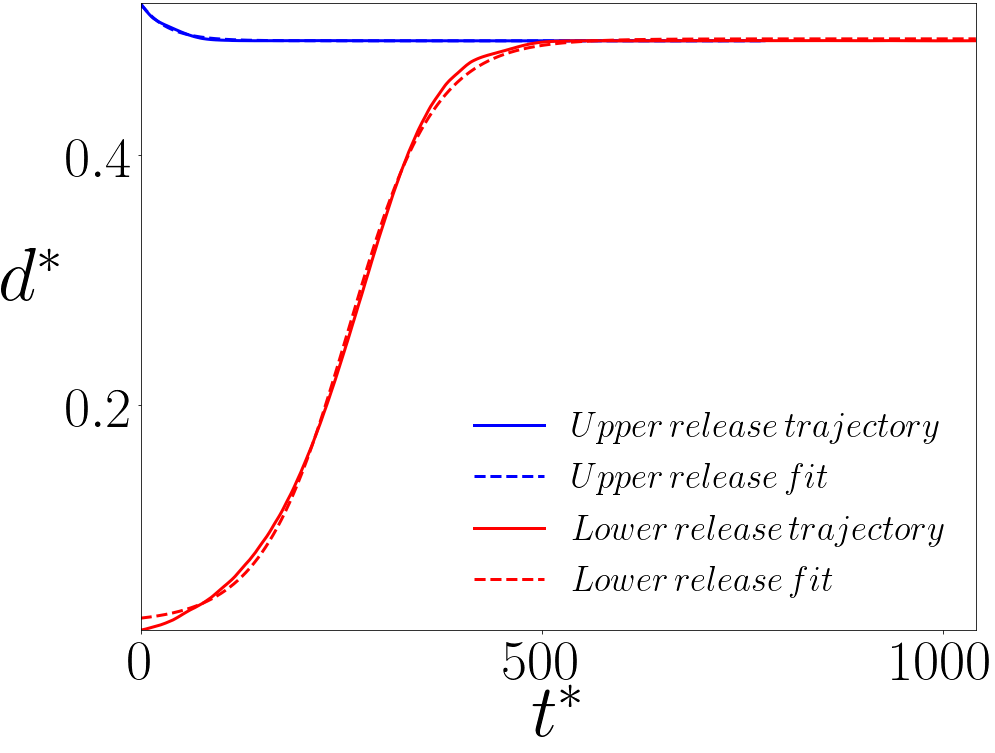}
   \caption{Regression fits to both steady trajectories at $Ca=0.09$}
   \label{Ca=0.09 regression fits}
\end{figure}

Taking the first derivative from the steady flow trajectory at the lowest $Ca$ ($Ca=0.09$) results in a very noisy curve that is impossible to interpret. Therefore, we use an accurate non-linear regression by minimizing the sum of squared errors to fit the trajectories with analytical expressions, from which we can take first and second derivatives analytically. The trajectory from the upper release is very similar to an exponential decay. Therefore, we fit a curve with the following form to it:
\begin{equation}\label{upper release trajectory}
 d(t) \approx ce^{bt} + k,
\end{equation}
Where $c$, $b$, and $k$ are all constants that should be determined after completing the optimization. The constant $k$ is essentially the drop equilibrium distance from the center. The trajectory from the lower release looks like the sigmoid logistic function. Consequently, we use the following equation as its analytical general form:
\begin{equation}\label{lower release trajectory}
 d(t) \approx \frac{c}{1+e^{-b(t-k)}} + offset,
\end{equation}
Where again, $c$, $b$, $k$, and $offset$ are the regressor constants. The regression fits to both trajectories from the upper and lower release have very high $R^2$ scores of 0.99 as plotted in fig. \ref{Ca=0.09 regression fits}. This figure further confirms that the drop focuses at the same $d^*$ regardless of its initial location.

Once the migration acceleration is derived following the aforementioned steps, it will be multiplied by $\frac{4}{3}\pi a^3 \rho$, which is the total constant mass of the drop with the initial spherical shape, to get the total force. By subtracting the calculated drag force from the total force, the total lift force can be obtained.
\section{\label{sec:level1}Results and Discussion\protect\\}
In this section, we report the results of a single deformable droplet simulations in the previously introduced microchannel that contains both steady and oscillatory carrier fluid. As we are interested in studying the effects of oscillation frequency and Capillary number on the lift force, we fix the $Re$ at a value of $10$. $Ca$ ranges between $0.09$ and $1.67$, and for oscillatory cases, $\omega^*$ values are chosen such that for a channel with a cross-section of $100 \mu m$ and water as the working fluid at room temperature, the frequency ranges between 2Hz and 200Hz, which is mostly referred to in the literature \cite{dincau2020pulsatile}. The validation of our numerical framework, as well as grid and domain independence studies, are discussed in detail in our previous work \cite{lafzi2021dynamics}.

\begin{figure}[h]
  \centering
  \includegraphics[width=5in]{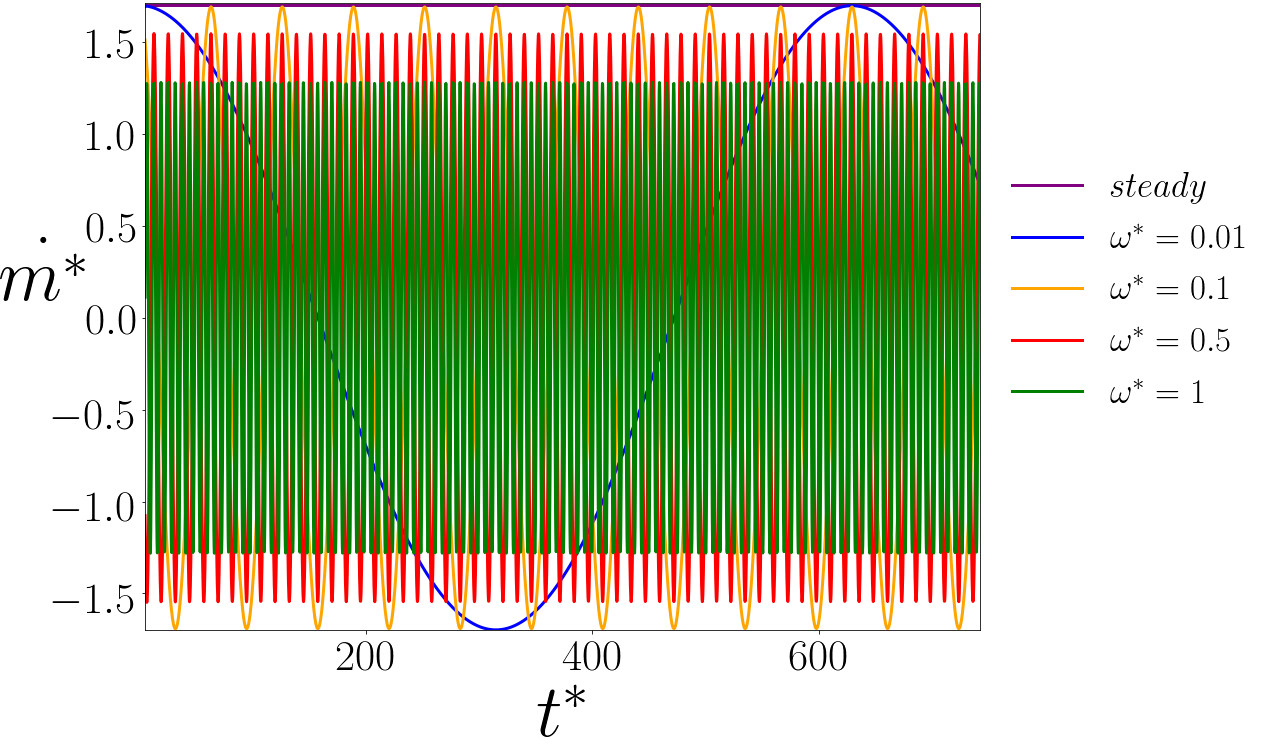}
   \caption{Flow rate versus time at $Ca=1.67$ and $Re=10$}
   \label{flow_rates}
\end{figure}

Figure \ref{flow_rates} illustrates the dimensionless mass flow rate over dimensionless time. It can be seen that while the steady regime has the largest constant flow rate in a single direction, the average of oscillatory flow rates in each half of a periodic cycle decreases by increasing the frequency \cite{lafzi2020inertial, lafzi2021dynamics}. Although the average of a sinusoidal function in half of a period is constant regardless of its oscillation frequency ($\frac{1}{\frac{\pi}{\omega}}\int_{0}^\frac{\pi}{\omega} sin({{\omega}t})dt=\frac{2}{\pi}$), the lower maximum absolute value of the flow rate at higher frequencies is the chief reason for the observed phenomenon.

\begin{figure}
\centering
  \begin{subfigure}[ht]{.488\textwidth}
  \centering
  \includegraphics[height=.225\textheight]{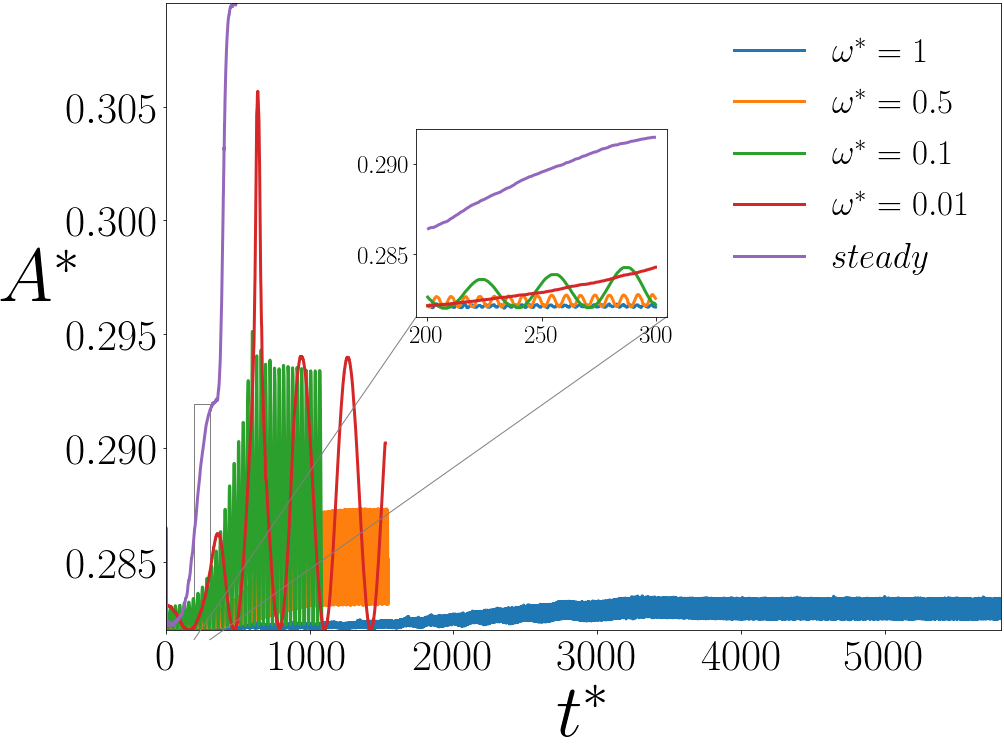}
  \caption{}
  \label{area_lower}
  \end{subfigure}
  ~
  \begin{subfigure}[ht]{.488\textwidth}
  \centering
  \includegraphics[height=.225\textheight]{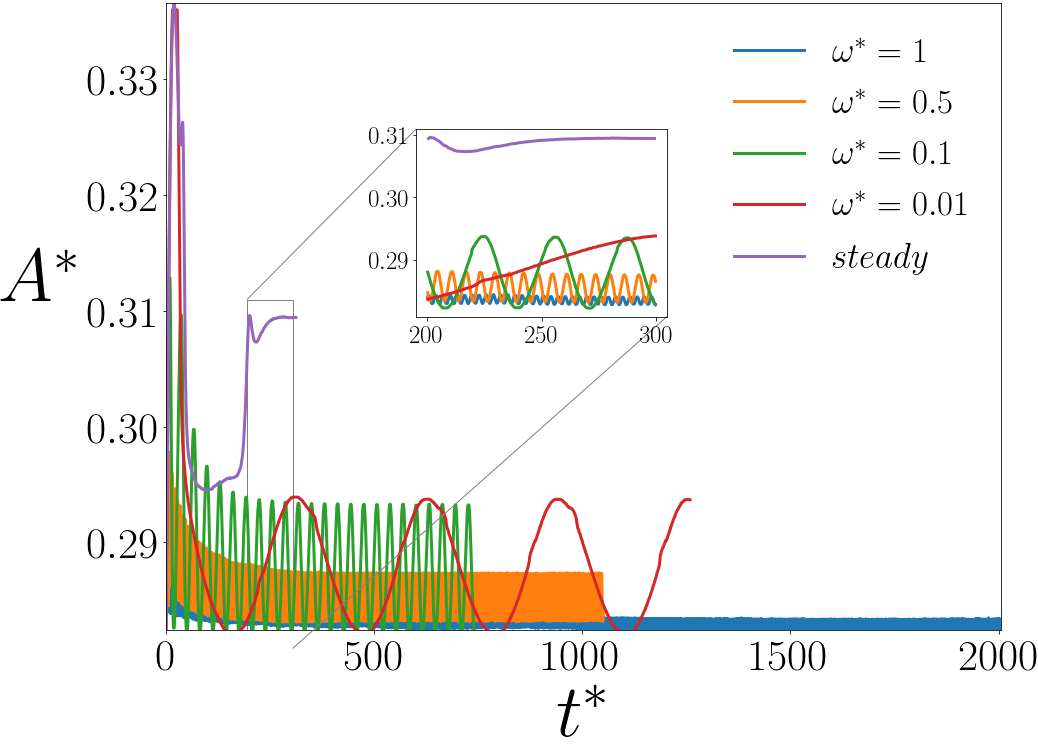}
  \caption{}
  \label{area_upper}
  \end{subfigure}
   \caption{Transient cross-stream frontal projected area of the drop when it is released from a (a) lower initial location, and an (d) upper initial location}
     \label{area_Ca=1.67}
\end{figure}

Figure \ref{area_Ca=1.67} visualizes the dimensionless time-dependent frontal projected area of the droplet (parameter $A_d$ in equation \ref{drag force}) as it migrates toward its lateral equilibrium position traveling both upper and lower-release trajectories. The first thing we note is that in the transient stage before focusing, the drop has a higher average projected area while traveling the upper trajectory (fig. \ref{area_upper}) compared to the one in the lower trajectory (fig. \ref{area_lower}) in each of the flow regimes correspondingly. This is because the drop experiences more shear and deforms easier when traveling the upper trajectory. Moreover, in each subfigure, the average of $A^*$ is lower at a higher frequency since the average deformation parameter decreases by increasing the frequency \cite{lafzi2021dynamics}. It is important to note that the minimum projected area of the drop in the steady flow is its initial value when the drop is still undeformed and has a spherical shape; in the oscillatory cases, this minimum value occurs when the direction of the flow changes in each periodic cycle. Also, as expected, the drop at higher $Ca$ deforms more and has a higher projected area. This is why $Ca=1.67$ is used for visualization here among all the other cases in the present study.

\begin{figure}[h]
  \centering
  \includegraphics[width=4.3in]{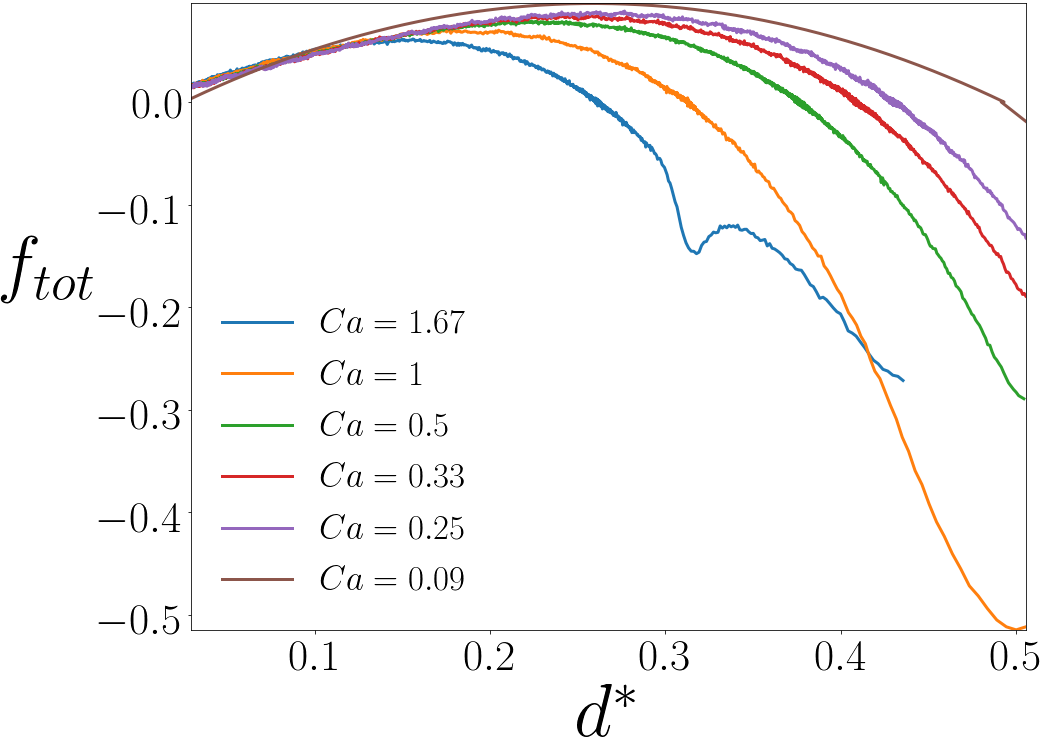}
   \caption{Total lift coefficient for the steady flows at different $Ca$}
   \label{all_steady_total_lift}
\end{figure}

Figure \ref{all_steady_total_lift} demonstrates the dimensionless total lift coefficient as a function of the dimensionless distance of the drop from the channel center in the steady flows and at different $Ca$. Similar to \cite{chen2014inertial}, all of the lift coefficients in this work are obtained by dividing the derived lift force, according to the introduced methodology in the previous section, by a factor of $\frac{\pi}{8}\rho{V^2_{avg,s}}(2a)^2$, in which $V_{avg,s}$ is the average of flow velocity across the channel cross-section in the steady flow. As expected, we observe that each lift curve has a stable equilibrium point at the corresponding drop focal point. Furthermore, at each $Ca$, the maximum positive total lift occurs when the drop migration velocity is also maximum. This maximum value is the highest at the lowest $Ca$. In addition, the maximum negative total lift is at the initial location of the upper trajectory, and its absolute value is the highest for higher $Ca$ except for $Ca=1.67$. This is because as we go further up from the channel center and the drop focal point, the deformation lift becomes the dominant force. According to equation \ref{deformation force}, this force is larger at higher $Ca$. Also, the negative lift sign in this region is due to the direction of the deformation-induced force, which is toward the center. The drop at $Ca=1.67$ is released from an initial location closer to the center compared to other cases because it has the highest deformability among all. When it was released from the same location as that of the others, it experienced an extremely large deformation that led to its break up. Therefore, the selected initial point for $Ca=1.67$ is the furthest possible one from the center that results in the largest possible deformation of the drop throughout its upper trajectory without its break up. Consequently, since the drop in this case starts to travel from a closer distance from the center, it has a lower maximum negative total lift compared to $Ca=1$ and $Ca=0.5$ (please refer to eq. \ref{deformation force} that shows the dependence of the deformation force on the drop distance from the center).   

\begin{figure}[h]
  \centering
  \includegraphics[width=4.3in]{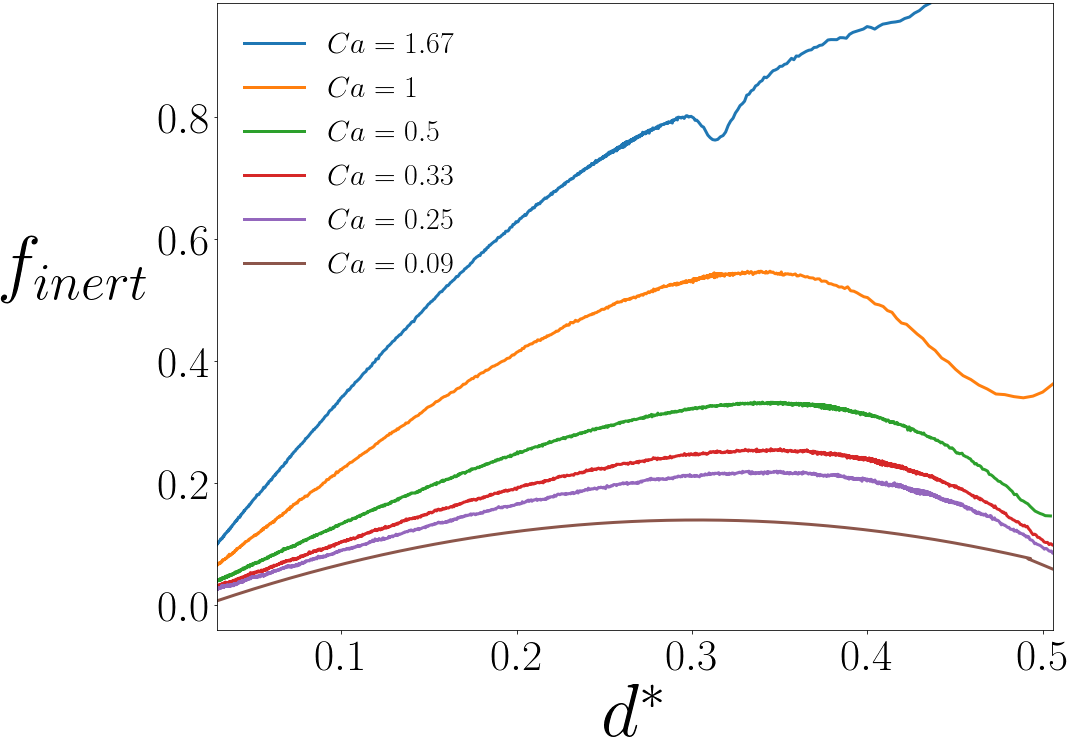}
   \caption{Inertial lift coefficient for the steady flows at different $Ca$}
   \label{all_steady_inertial_lift}
\end{figure}

Since the principal hypothesis underlying equation \ref{deformation force} is that the wall effect is negligible due to the large distance of the drop from it \cite{zhang2016fundamentals, ho1974inertial, jahromi2019improved}, we can assume that the shear-gradient force is the dominant component of the inertial lift in our study. After subtracting the calculated deformation lift based on this equation from the obtained total lift (fig. \ref{all_steady_total_lift}), we can derive the inertial lift, as shown in fig. \ref{all_steady_inertial_lift}. We see that the inertial lift coefficient increases as we increase the $Ca$ number. This could make sense as the more deformed shape of the drop can help further increase the difference between the relative velocities of the fluid with respect to the drop on the channel wall and center sides, which is the chief reason for the shear-gradient force existence \cite{zhang2016fundamentals}. According to eq. \ref{deformation force}, the deformation force is a linear function of the drop distance from the center and is larger for higher values of $Ca$. Because of this trivial conclusion, a plot of this force is not depicted here.

\begin{figure}
\centering
  \begin{subfigure}[ht]{.488\textwidth}
  \centering
  \includegraphics[height=.225\textheight]{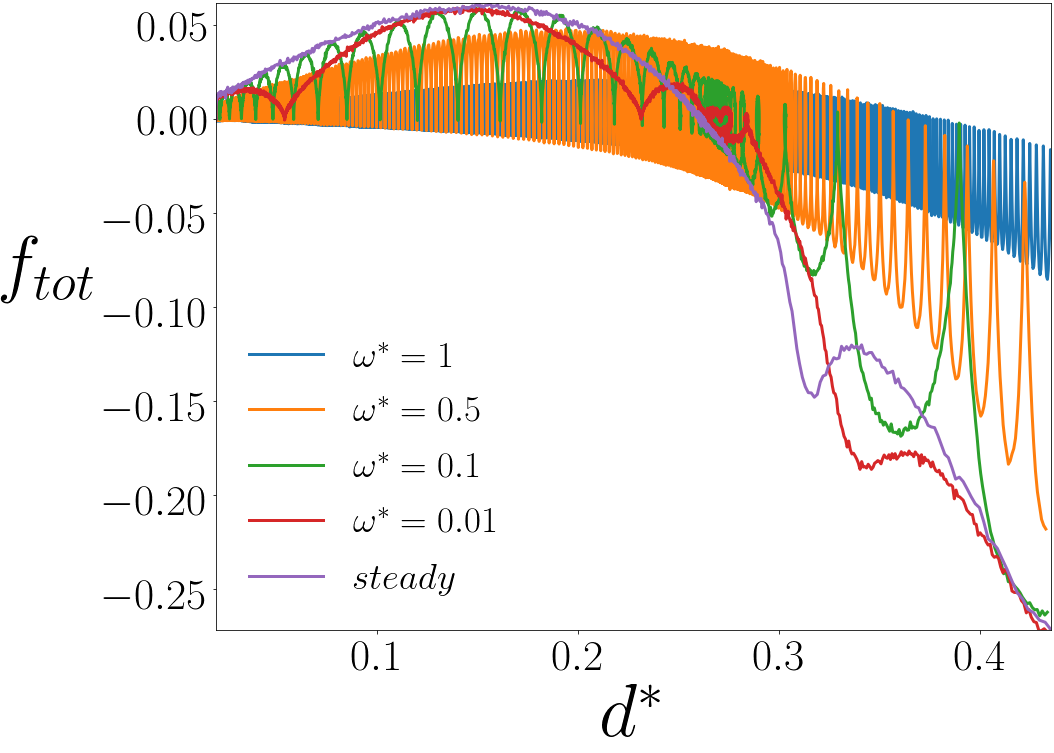}
  \caption{}
  \label{all_total_lift_Ca=167}
  \end{subfigure}
  ~
  \begin{subfigure}[ht]{.488\textwidth}
  \centering
  \includegraphics[height=.225\textheight]{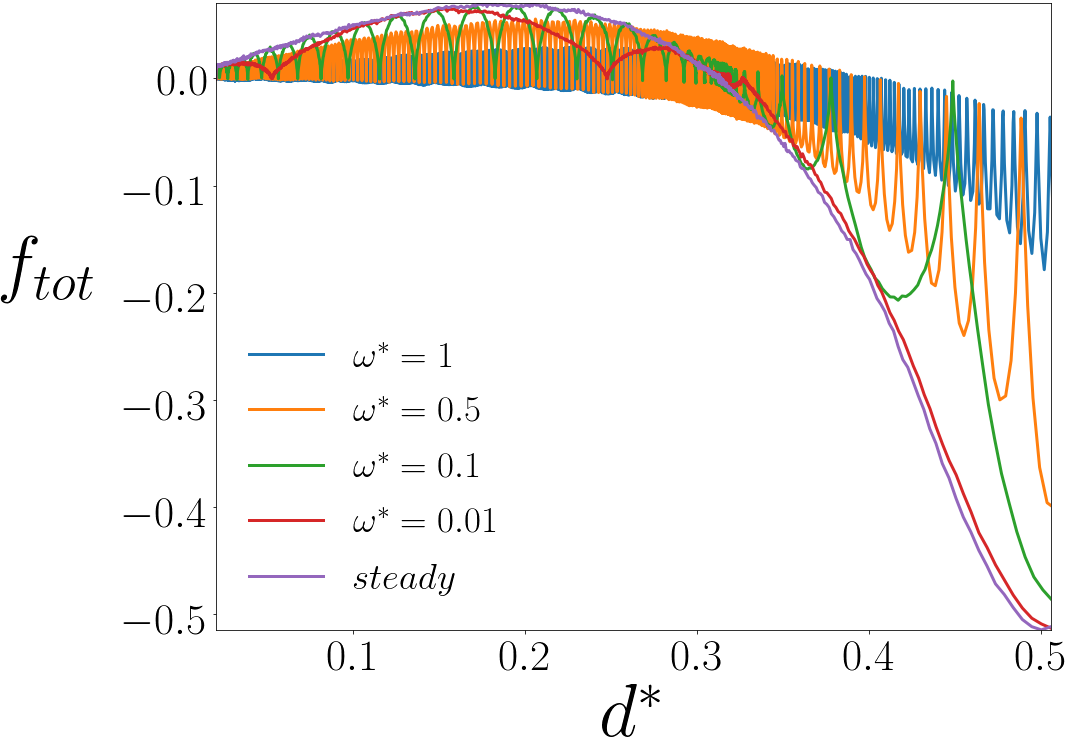}
  \caption{}
  \label{all_total_lift_Ca=1}
  \end{subfigure}
  ~
  \begin{subfigure}[ht]{.488\textwidth}
  \centering
  \includegraphics[height=.225\textheight]{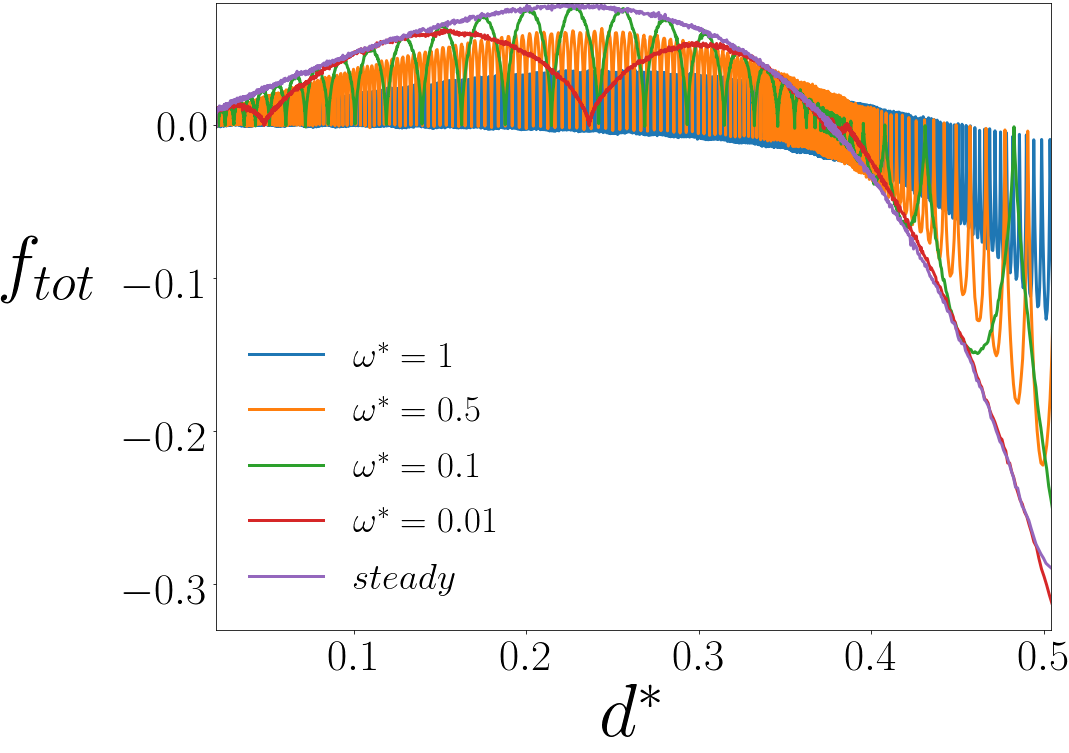}
  \caption{}
  \label{all_total_lift_Ca=0.5}
  \end{subfigure}
  ~
  \begin{subfigure}[ht]{.488\textwidth}
  \centering
  \includegraphics[height=.225\textheight]{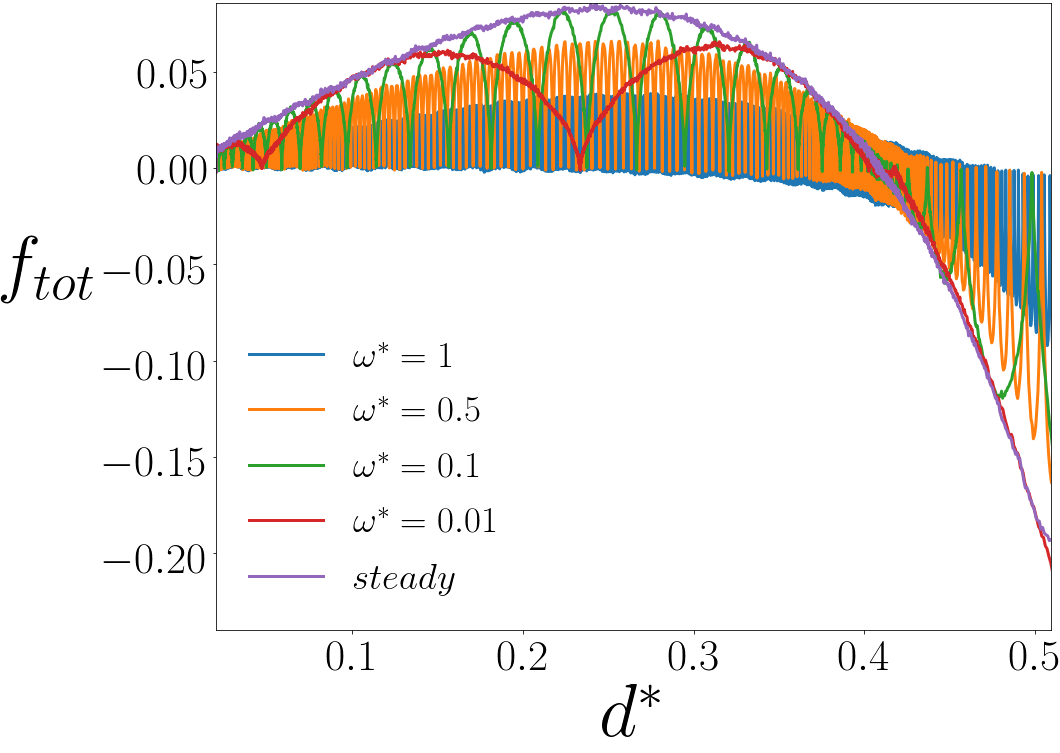}
  \caption{}
  \label{all_total_lift_Ca=0.33}
  \end{subfigure}
  ~
  \begin{subfigure}[ht]{.488\textwidth}
  \centering
  \includegraphics[height=.225\textheight]{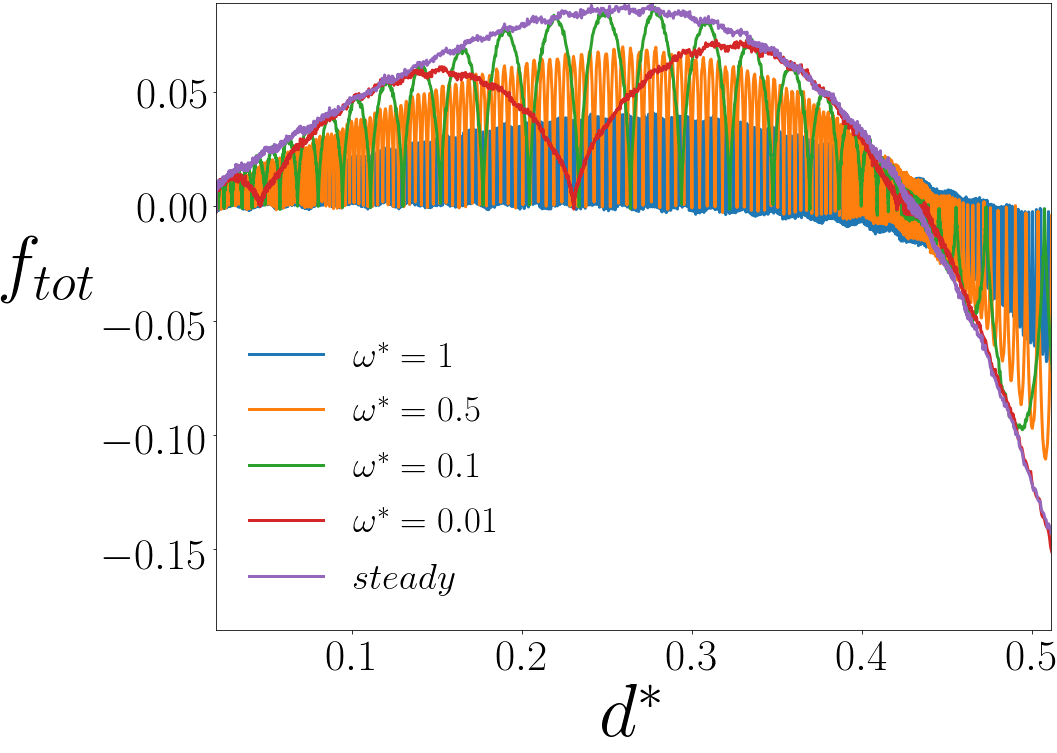}
  \caption{}
  \label{all_total_lift_Ca=0.25}
  \end{subfigure}
  ~
   \caption{Total lift coefficients for steady and oscillatory flows with different frequencies at (a) $Ca=1.67$, (b) $Ca=1$, (c) $Ca=0.5$, (d) $Ca=0.33$, and (e) $Ca=0.25$}
     \label{all_total_lift}
\end{figure}

Total lift curves acting on the drop in steady and different oscillatory flows at a few $Ca$ numbers are expressed in fig. \ref{all_total_lift}. In each subfigure, the higher the drop migration velocity, the larger are both the amplitude of oscillations and the distance between two corresponding points (e.g. maximum or minimum in the oscillatory cycle) on two consecutive periodic cycles. Hence, similar to steady regimes, the maximum absolute values of oscillatory lift coefficients occur when the drop migration velocities are maximum as well. Similarly, the lift oscillations around the drop focal point and near the lower initial point are lower because the drop migration velocities are minimum at those locations. 

\begin{figure}
\centering
  \begin{subfigure}[ht]{.488\textwidth}
  \centering
  \includegraphics[height=.225\textheight]{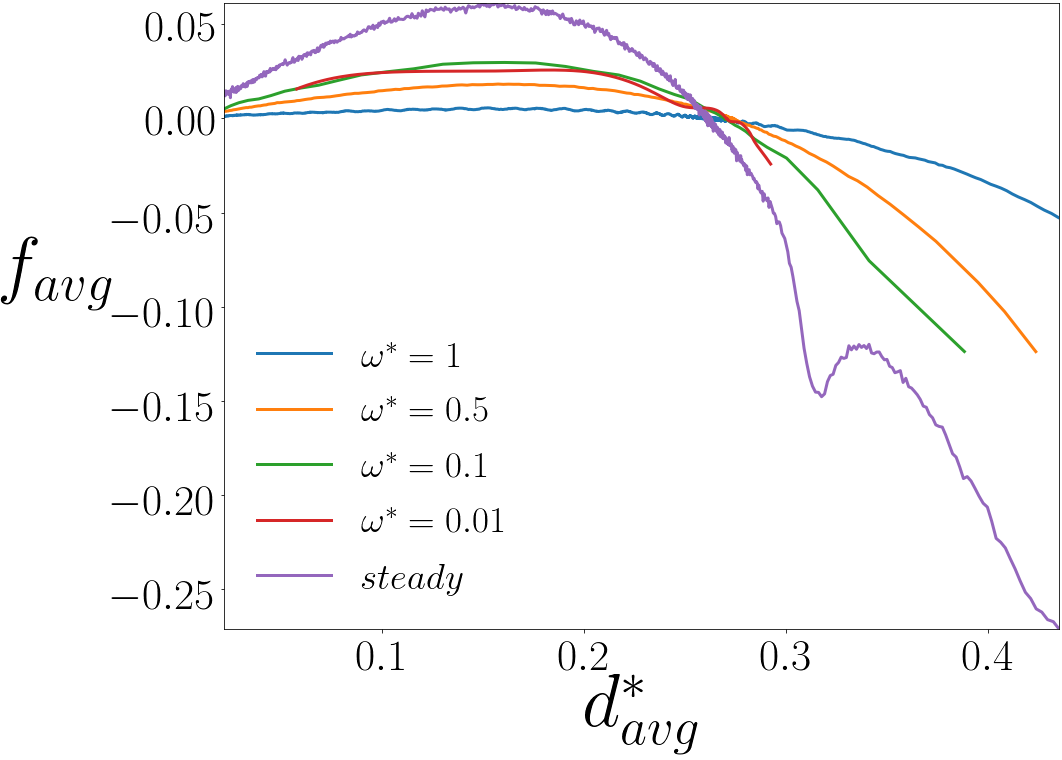}
  \caption{}
  \label{all_averaged_lift_Ca=167}
  \end{subfigure}
  ~
  \begin{subfigure}[ht]{.488\textwidth}
  \centering
  \includegraphics[height=.225\textheight]{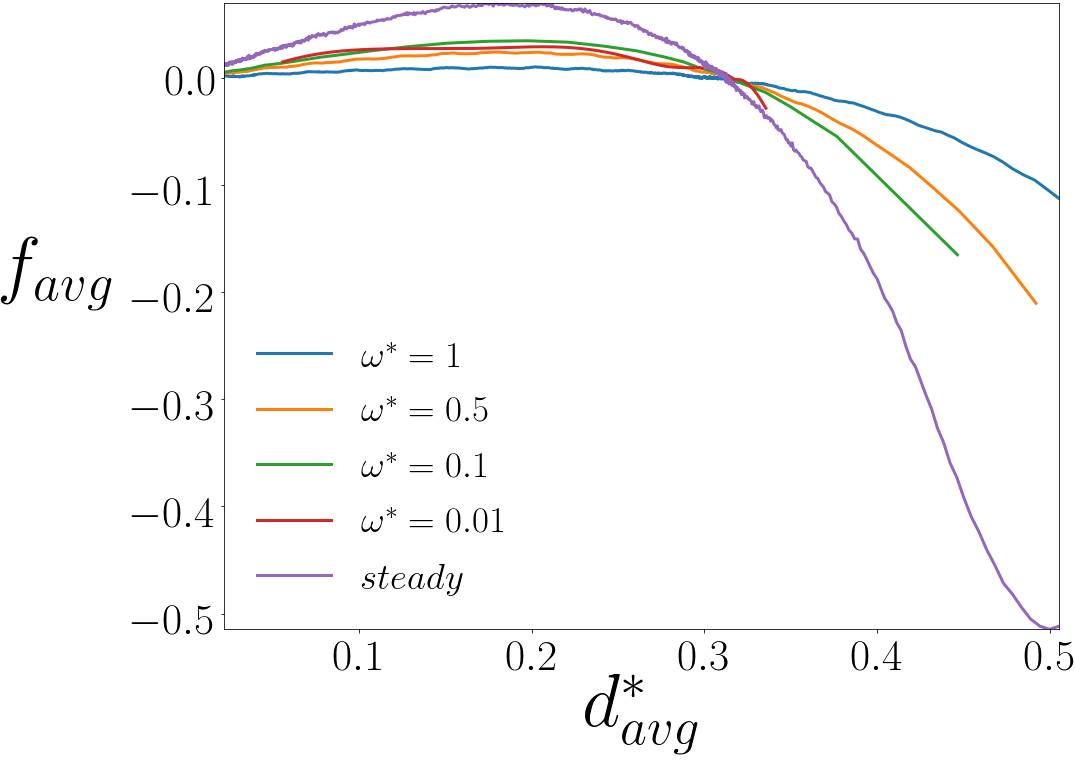}
  \caption{}
  \label{all_averaged_lift_Ca=1}
  \end{subfigure}
  ~
  \begin{subfigure}[ht]{.488\textwidth}
  \centering
  \includegraphics[height=.225\textheight]{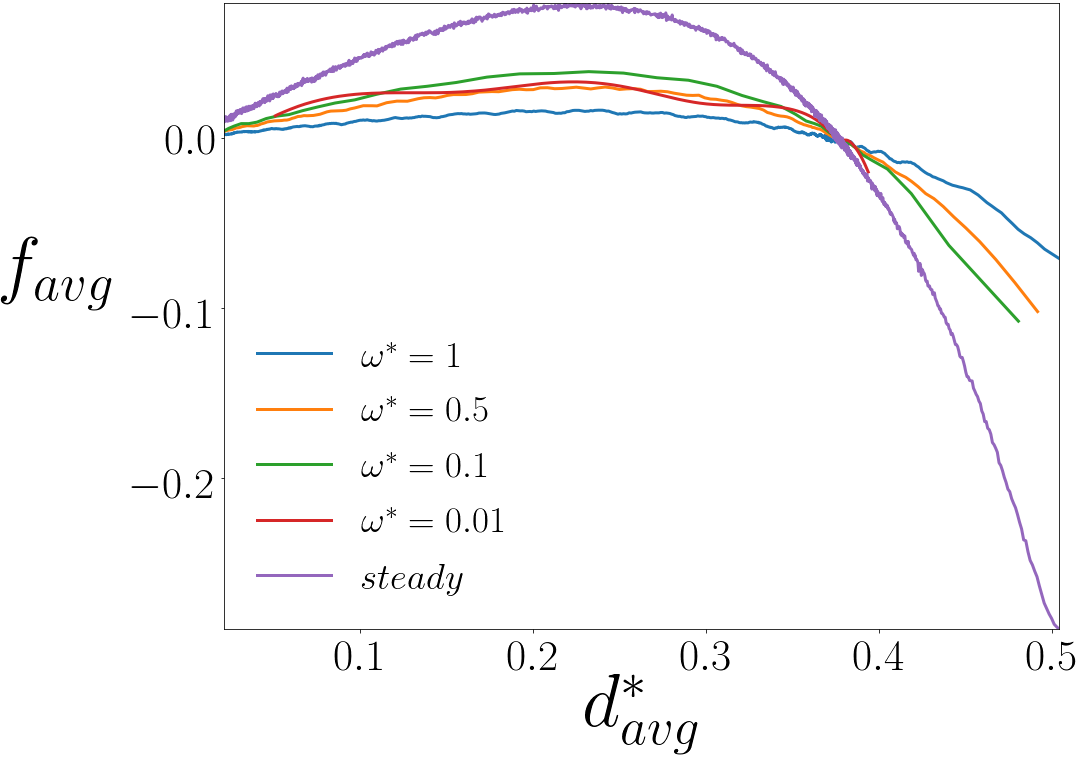}
  \caption{}
  \label{all_averaged_lift_Ca=0.5}
  \end{subfigure}
  ~
  \begin{subfigure}[ht]{.488\textwidth}
  \centering
  \includegraphics[height=.225\textheight]{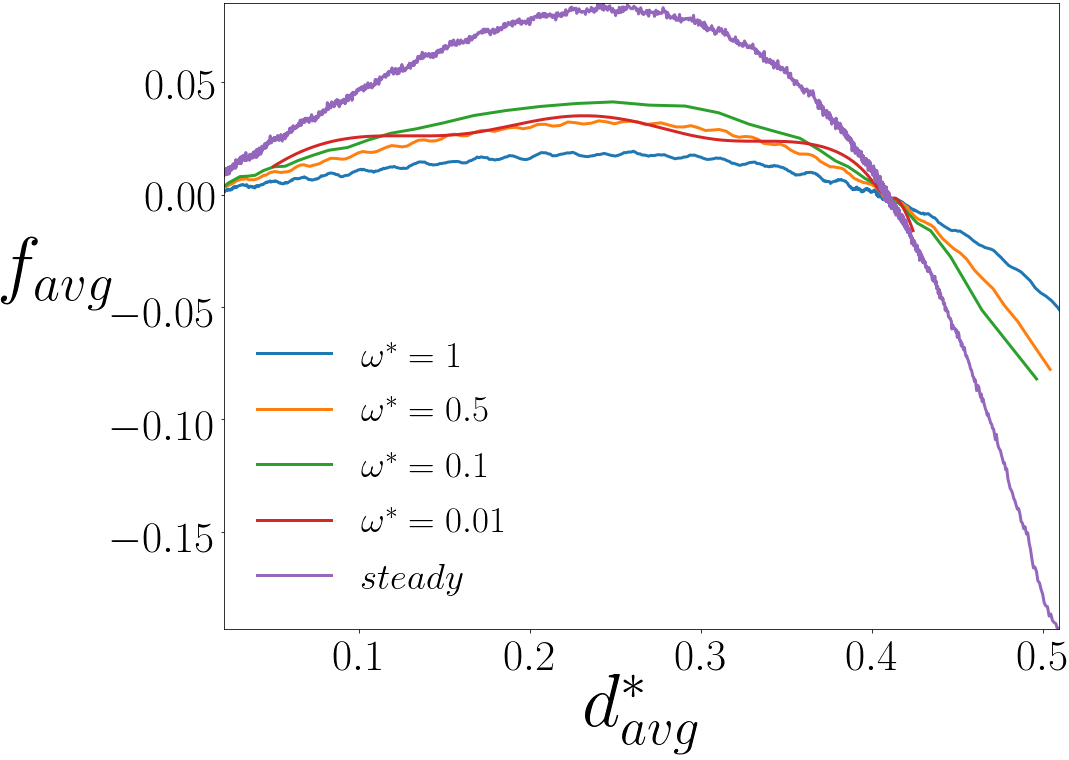}
  \caption{}
  \label{all_averaged_lift_Ca=0.33}
  \end{subfigure}
  ~
  \begin{subfigure}[ht]{.488\textwidth}
  \centering
  \includegraphics[height=.225\textheight]{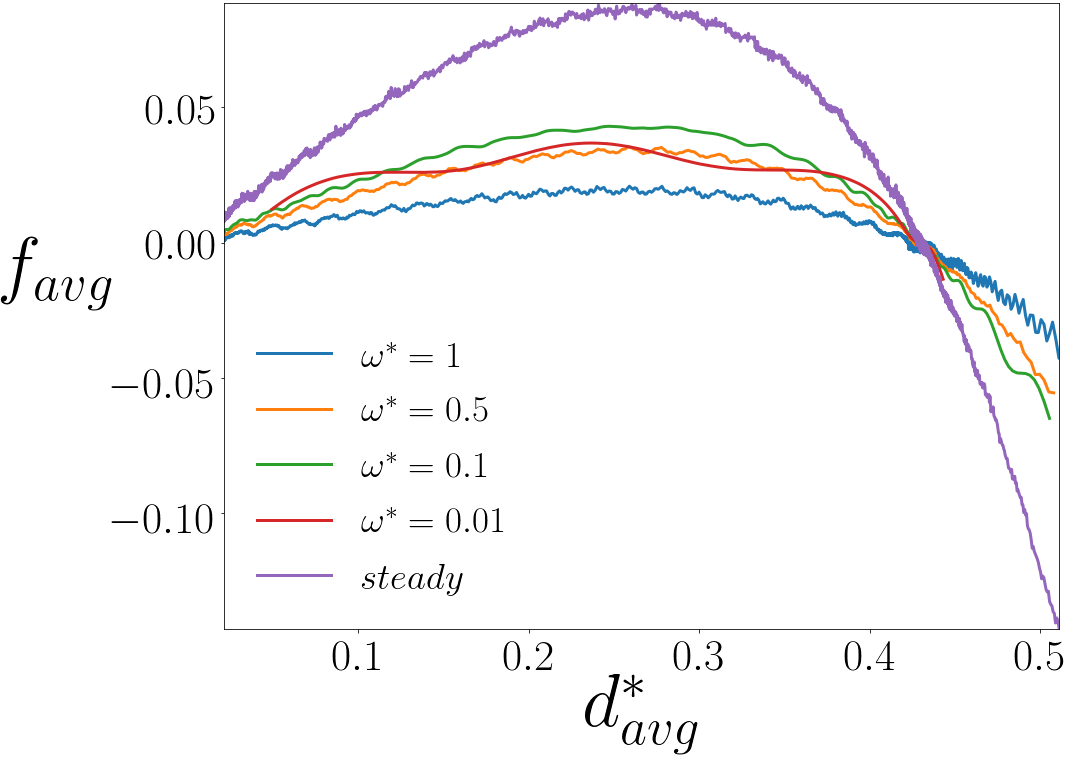}
  \caption{}
  \label{all_averaged_lift_Ca=0.25}
  \end{subfigure}
  ~
   \caption{Averaged total lift coefficients for steady and oscillatory flows with different frequencies at (a) $Ca=1.67$, (b) $Ca=1$, (c) $Ca=0.5$, (d) $Ca=0.33$, and (e) $Ca=0.25$}
     \label{all_averaged_lift}
\end{figure}

The moving averages of the total lift coefficients in fig. \ref{all_total_lift} are plotted in fig. \ref{all_averaged_lift}. The selected $d^*_{avg}$ while computing the average of lift in each corresponding oscillatory cycle is chosen to be the middle value of $d^*$ in that period. Therefore, since oscillatory cycles with lower frequencies have longer periods, the averaged lift curves at lower frequencies cover a shorter length of $d^*_{avg}$ (a later beginning and a sooner ending). We first note that the obtained averaged lift curves for the oscillatory flows are not necessarily as smooth as that of the steady lift at the corresponding $Ca$. This observation becomes more pronounced as we increase $\omega^*$ or decrease $Ca$. Nevertheless, these fluctuations on the curves are negligible compared to those of the original oscillatory lifts (fig. \ref{all_total_lift}). Additionally, although the average of inertial and deformation-induced lift forces decrease separately by increasing $\omega^*$ \cite{lafzi2021dynamics}, the difference between them (fig. \ref{all_averaged_lift}) does not follow the same pattern. This confirms the existence of a drop focal point with an extremum distance from the channel center at an intermediate frequency, as elaborated in our previous work \cite{lafzi2021dynamics}. Despite this, the average of lift is the largest in the steady flow and the smallest in the oscillatory flow with the highest frequency at each $Ca$ in our study.

\begin{figure}
\centering
  \begin{subfigure}[ht]{.488\textwidth}
  \centering
  \includegraphics[height=.225\textheight]{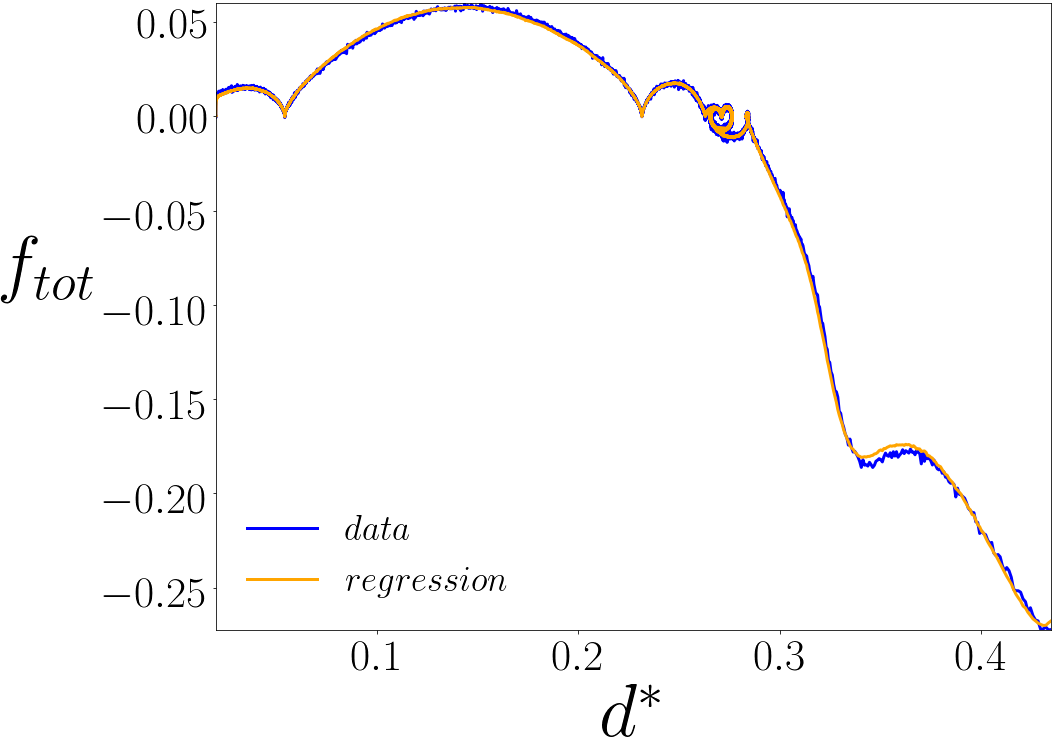}
  \caption{}
  \label{lift_regression_Ca=167_w=0.01}
  \end{subfigure}
  ~
  \begin{subfigure}[ht]{.488\textwidth}
  \centering
  \includegraphics[height=.225\textheight]{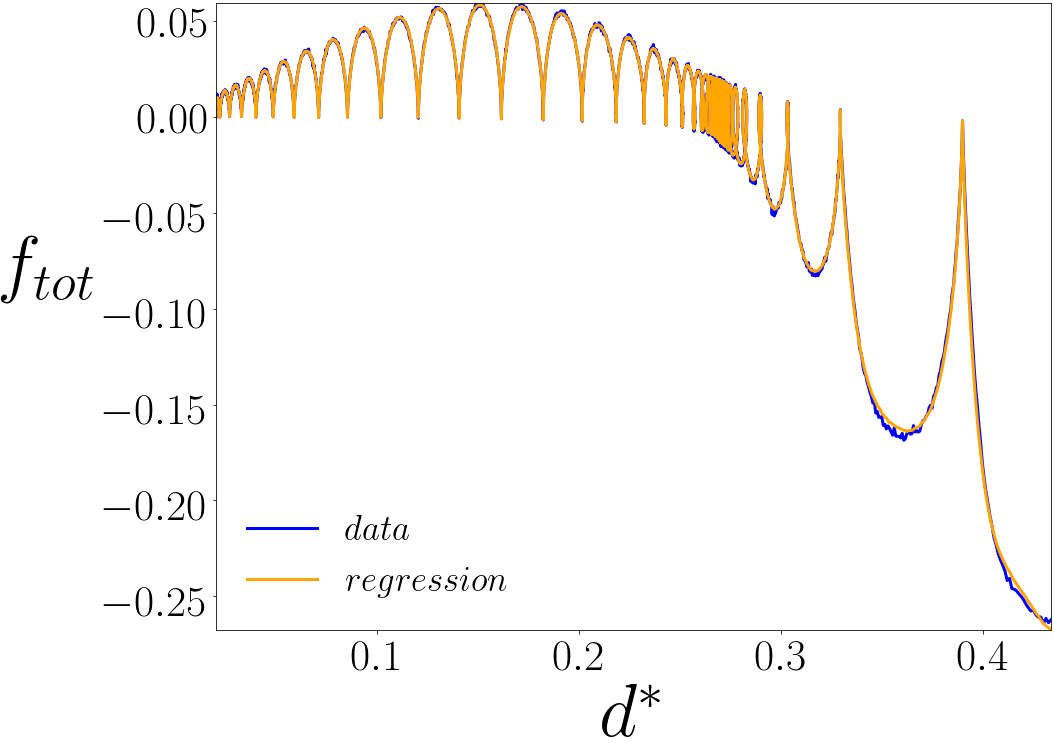}
  \caption{}
  \label{lift_regression_Ca=167_w=0.1}
  \end{subfigure}
  ~
  \begin{subfigure}[ht]{.488\textwidth}
  \centering
  \includegraphics[height=.225\textheight]{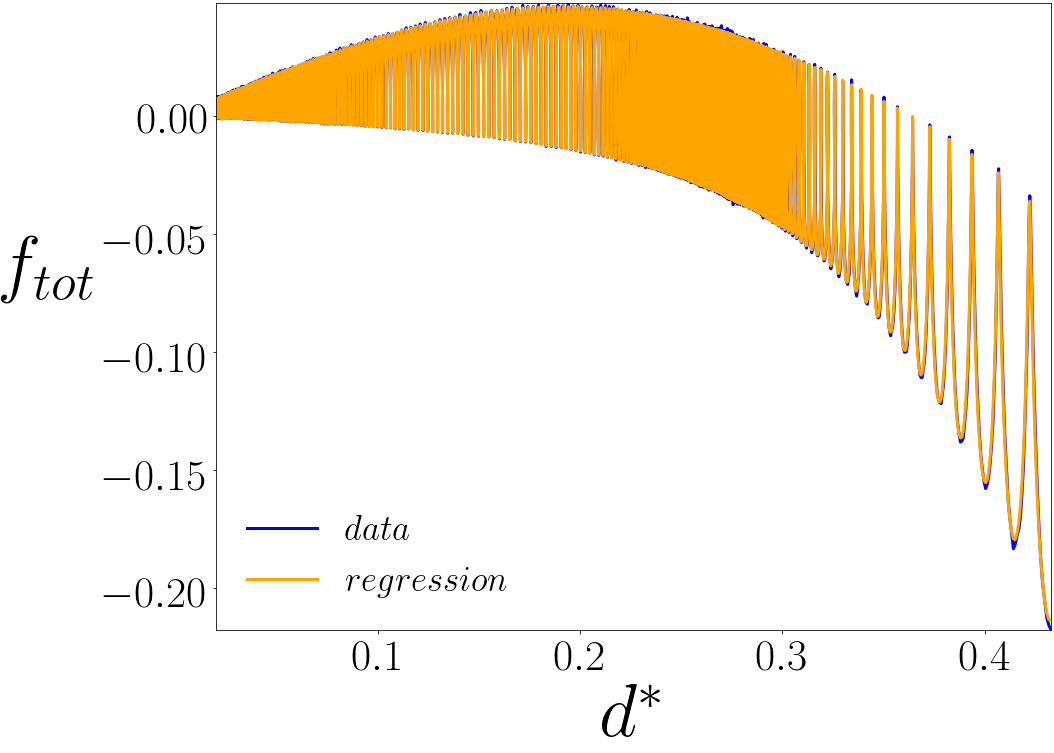}
  \caption{}
  \label{lift_regression_Ca=167_w=0.5}
  \end{subfigure}
  ~
  \begin{subfigure}[ht]{.488\textwidth}
  \centering
  \includegraphics[height=.225\textheight]{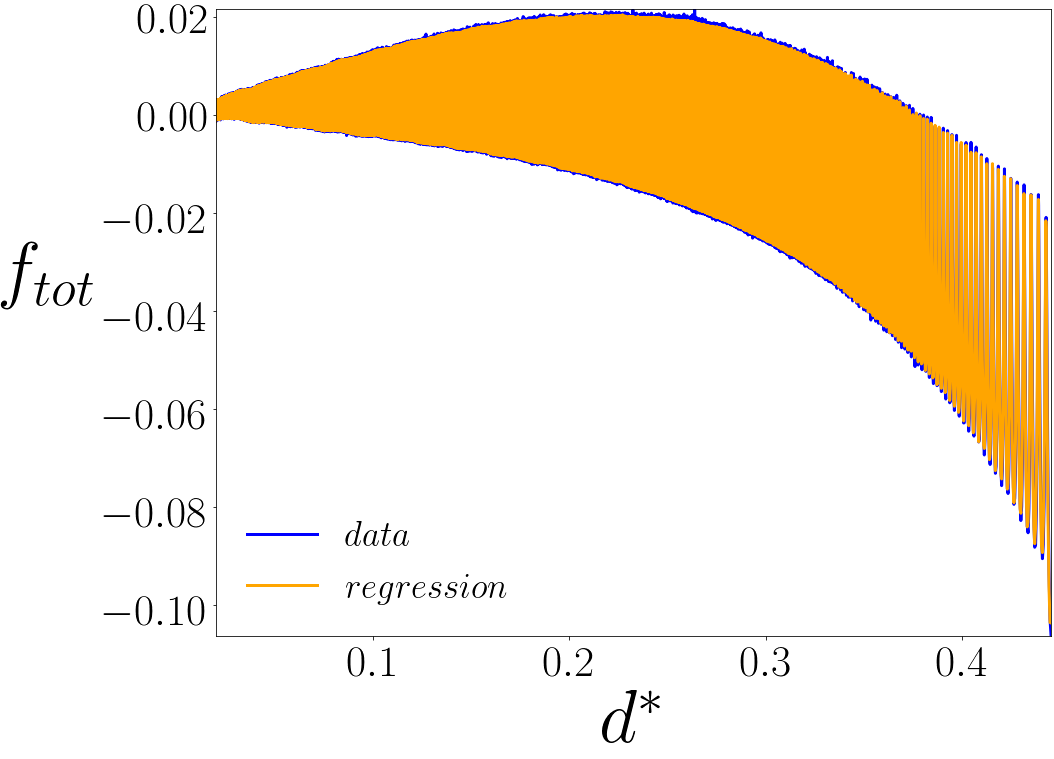}
  \caption{}
  \label{lift_regression_Ca=167_w=1}
  \end{subfigure}
   \caption{The fitted total lift coefficients to the ground-truth data using the introduced non-linear regression at $Ca=1.67$ and (a) $\omega^*=0.01$, (b) $\omega^*=0.1$, (c) $\omega^*=0.5$, and (d) $\omega^*=1$}
     \label{all_lift_regression_Ca=167}
\end{figure}

By taking another close look at fig. \ref{all_total_lift}, we realize that each of the oscillatory total lift coefficients can be fitted using an expression that comprises of a base curve, which can be best fitted by a 4th order polynomial, combined with the absolute value of some sinusoidal oscillations. Both the amplitude of oscillations and oscillatory periods can be controlled by the drop migration velocity and have direct relationship with it. In other words, the proposed expression can have the following form:
\begin{equation}\label{regression_curve}
f_{tot}\approx\left\{m+nv(t)\left|cos\left(\frac{c}{av(t)}d(t)+b\right)\right|\right\}\left\{gd^4(t)+hd^3(t)+kd^2(t)+ld(t)+q\right\}
\end{equation}
Where $d(t)$ and $v(t)$ are the time-dependent drop distance from the channel center and its migration velocity, respectively, and $m$, $n$, $a$, $b$, $c$, $g$, $h$, $k$, $l$, and $q$ are the constants to be determined while performing the optimization. The constant $m$ is just placed for achieving higher accuracy for the fits, and it ends up to be almost zero compared to other constants in the expression. The resultant curves are plotted in fig. \ref{all_lift_regression_Ca=167} against their corresponding data (fig. \ref{all_total_lift_Ca=167}) for $Ca=1.67$ and $\omega^*=0.01$, $\omega^*=0.1$, $\omega^*=0.5$, and $\omega^*=1$ with $R^2$ scores of 0.99, 0.99, 0.99, and 0.97, respectively. Similar curves having the same proposed expression with $R^2$ scores of 0.97 or higher and capable of capturing all the infinitesimal details are obtained for other $Ca$ numbers as well.

To further extend the total lift prediction to more general cases within a continuous range of $Ca$ and $\omega^*$, we consider the steady and averaged oscillatory lifts (fig. \ref{all_averaged_lift}) for regression with a 4th order polynomial here. In other words, the expression in the first bracket of eq. \ref{regression_curve} is replaced with a value of 1. Using the analytical set of equations \ref{equations_for_coeffients_determination}, we can derive the unknown coefficients $a$, $b$, $c$, $g$, and $h$ of the polynomial analytically. In these equations, the subscripts for $d$ and $f$ denote their locations. For instance, the subscript max represents where the magnitude of total (or averaged) lift is maximum and its first derivative is zero. The obtained polynomials with this approach have $R^2$ scores of around 0.9 or higher for the cases presented in fig. \ref{all_averaged_lift}.
\begin{equation}\label{equations_for_coeffients_determination}
\begin{cases}
     ad_{first}^4+bd_{first}^3+cd_{first}^2+gd_{first}+h=f_{first},\\
      ad_{max}^4+bd_{max}^3+cd_{max}^2+gd_{max}+h=f_{max},\\
      4ad_{max}^3+3bd_{max}^2+2cd_{max}+g=0,\\
      ad_{eq}^4+bd_{eq}^3+cd_{eq}^2+gd_{eq}+h=0,\\
      ad_{last}^4+bd_{last}^3+cd_{last}^2+gd_{last}+h=f_{last},
\end{cases}  
\end{equation}
Parameters $d_{first}$, $d_{max}$, $d_{eq}$, $d_{last}$, $f_{first}$, $f_{max}$, and $f_{last}$ are already available for the cases in fig. \ref{all_averaged_lift} to solve the system of equations \ref{equations_for_coeffients_determination} for them. However, we use the multi-fidelity Gaussian processes (MFGP) method to predict these unknown parameters for any given double inputs of $0.25\leq Ca\leq1.67$ and $0\leq\omega^*\leq1$. MFGP is a Bayesian stochastic approach that does a casual inference on a set of high and low-fidelity datasets, and it is extremely effective if there are strong correlations between them \cite{perdikaris2017nonlinear}. This method is described in detail in our previous work, and it is carried out to predict the distance of the drop equilibrium position from the channel center $(d_{eq})$ with $R^2$ of 0.99 and root mean squared error (RMSE) of 0.01 in that work \cite{lafzi2021dynamics}. Here, we refer to the data for the cases in fig. \ref{all_averaged_lift} except for $Ca=1$ as our high-fidelity data. We generate similar data for all the cases in that figure, but with a grid of $128\times 76 \times 76$ in the $x$, $y$, and $z$ directions, respectively, and having 13038 triangular elements for the discretization of the drop. We consider this data as our low-fidelity dataset. Therefore, we have a total of 25 low and 20 high-fidelity data points, which satisfies the required nested structure to apply MFGP on the data \cite{perdikaris2017nonlinear}. We randomly allocate 5 data points of the entire high-fidelity dataset as our test set since the high-fidelity response is our main target. We train the algorithm on the remaining 40 training data points and evaluate its performance on the test set. We repeat this procedure 30 times and compute the average of evaluation metrics so that the selection of test sets does not significantly affect the overall algorithm performance. 

Table \ref{MFGP evaluation} presents the average of $R^2$ and RMSE on our 6 remaining unknown parameters after completing the aforementioned steps. We can see that the trained algorithm is capable of predicting the intended parameters with very high accuracies. Especially, the accurate prediction of $f_{max}$ and $f_{last}$ is useful for determining the maximum and minimum values of averaged total lift for any given input in the range, respectively. The slightly less accurate prediction for $d_{max}$ (i.e. where the maximum averaged lift occurs) is because of the present randomness in its values among different cases. This is unlike the consistent pattern that exists for other parameters for a combination of $\omega^*$ values across different $Ca$ numbers. The similar lower accurate prediction for $d_{last}$ is due to the lack of data points between $Ca=1$ and $Ca=1.67$ since all the cases with $Ca\leq1$ have the same upper release initial location. However, the $R^2$ score of around 0.8 for this prediction is still high, and it can help us determine the furthest starting point from the center and the widest traveling region of a droplet with $1\leq Ca \leq1.67$ in the microchannel so that it undergoes the largest possible deformation without breaking up.

\begin{table}[h]
   \centering
   \begin{tabular}{|P{4.12cm}|P{4.12cm}|P{4.12cm}|}
     \hline
     \textbf{Parameter} & \boldsymbol{$R^2$}  & \textbf{RMSE} \\ \hline
     $d_{max}$ & 0.79 & 0.0140\\ \hline
    $d_{last}$ & 0.78 & 0.0213 \\ \hline
      $d_{first}$ & 0.97 & 0.0007 \\ \hline
      $f_{last}$ & 0.92 & 0.0177 \\ \hline
      $f_{max}$ & 0.99 & 0.0015 \\ \hline
      $f_{first}$ & 0.99 & 0.0003 \\ 
     \hline
   \end{tabular}
   \caption{MFGP averaged performance metrics on 30 randomly chosen test sets for the parameters required for the determination of the analytical averaged total lift polynomial coefficients}
   \label{MFGP evaluation}
\end{table}

\section{Conclusions}
The dynamics of particles and biological cells in microchannels has caught many researchers' attention because of several biomicrofluidic applications it has. The underlying physics owes its behavior mainly to the presence of different lift forces in such channels. Hence, many scientists have dedicated their time to calculate or measure these forces. However, most of these works have focused on analyzing the lift forces acting on solid and non-deformable particles and studied the effects of parameters such as particles' size, Reynolds number, etc on them. Consequently, such analysis on deformable droplets or bubbles and studying the effects of varying parameters like Capillary number is almost missing in the literature. In this work, we have extended such analysis to the case of a single deformable droplet in the channel. We have calculated the main components of the lift force based on a unique methodology that merely depends on the drop trajectory. To do so, first, the drop migration velocity and its frontal projected area as it travels its lateral trajectory have been computed to calculate the drag force in the wall-normal direction accurately. After applying Newton's second law on the drop, the total lift profile is obtained over a region where the drop has a distance higher than its diameter from the wall. It has been observed that the total lift has a higher maximum at a lower Capillary, and its minimum decreases as we increase the $Ca$. The inertial and deformation-induced lift forces both increase by increasing the $Ca$ number. Moreover, since the oscillatory flows within the microchannel were previously shown to enable working with sub-micron biological particles as well as introducing new focal points for them, we have also included these flow regimes in our analysis and investigated the effects of oscillation frequency on the lift in addition to the Capillary number. We have seen that for all cases, the total lift and for oscillatory ones, the amplitude of oscillations are both higher when the drop migration velocity is higher. At each $Ca$, the steady lift and moving averages of oscillatory ones at different $\omega^*$ have also been compared. It has been shown that the steady lift has the largest magnitude, and the average of oscillatory one with the highest frequency in this study has the smallest strength. However, there is not a constant decreasing pattern in the average of lift by increasing the frequency, which is why the drop focuses furthest from the channel center at an intermediate $\omega^*$. Additionally, an accurate mathematical expression has been proposed that captures the detailed total oscillatory lift curves at various $\omega^*$ with $R^2$ scores of 0.97 or higher. Finally, the multi-fidelity Gaussian processes has been used to accurately predict the 7 unknown parameters required to define a simple 4th order polynomial to fit the steady and averaged oscillatory lifts with $R^2$ scores of about 0.9 or higher for any given $Ca$ and $\omega^*$ within the ranges of $0.25\leq Ca\leq1.67$ and $0\leq\omega^*\leq1$. 
\bibliography{sorsamp}

\end{document}